\documentclass[aip,graphicx,preprint]{revtex4-1}

\draft % marks overfull lines with a black rule on the right
\usepackage{subfigure}
\usepackage{graphicx}
\usepackage{here}
\usepackage{bm}
\usepackage{amsmath}
\usepackage{amssymb}
\usepackage{xcolor}
\usepackage[normalem]{ulem}
\usepackage{url}

\newcommand{\mb}{\mathbf}
\newtheorem{theorem}{Theorem}[section]

\newtheorem{remark}[theorem]{Remark}

\definecolor{dgn}{rgb}{0,0.7, 0.2}

\definecolor{orange}{rgb}{1, .5, 0.0}

\begin{document}

\preprint{}

\title{Long-term prediction of El Ni\~{n}o-Southern Oscillation using reservoir computing with data-driven realtime filter}


\author{Takuya Jinno}
\affiliation{Faculty of Sustainable Design, University of Toyama, Toyama 930-8555, Japan}
\author{Takahito Mitsui}
\affiliation{Faculty of Health Data Science, Juntendo Univerity, Urayasu, Chiba 279-0013, Japan}
\author{Kengo Nakai}
\affiliation{The Graduate School of Environment, Life, Natural Science and Technology, Okayama University, Okayama 700-8530, Japan}
\author{Yoshitaka Saiki}
\affiliation{Graduate School of Business Administration, Hitotsubashi University, Tokyo 186-8601, Japan}
\author{Tsuyoshi Yoneda}
\affiliation{Graduate School of Economics, 
Hitotsubashi University, Tokyo 186-8601, Japan}

\date{\today}

\begin{abstract}
In recent years, the application of machine learning approaches to time-series forecasting of climate dynamical phenomena has become increasingly active. It is known that applying a band-pass filter to a time-series data is a key to obtaining a high-quality data-driven model. Here, to obtain longer-term predictability of machine learning models, we introduce a new type of band-pass filter. It can be applied to realtime operational prediction workflows since it relies solely on past time series. We combine the filter with reservoir computing, which is a machine-learning technique that employs a data-driven dynamical system. As an application, we predict the multi-year dynamics of the El Ni\~{n}o-Southern Oscillation with the prediction horizon of 24 months using only past time series.
\end{abstract}

\maketitle %\maketitle must follow title, authors, abstract and \pacs

\begin{quotation}
%\textcolor{red}{
\textit{Lead paragraph:}\quad 
Predicting long-term climate phenomena, such as the El Ni\~{n}o–Southern Oscillation (ENSO), is crucial for mitigating the impacts of extreme weather events worldwide on agriculture and economics. Yet, forecasting the complex systems remains a challenge, especially when different time scales of variability overlap. Furthermore, in past studies which use conventional smoothing technique, information from the future is introduced to some extent. 
%In this paper, we introduce a new type of filter that is applicable to realtime operational forecast and combine it with a machine-learning framework known as reservoir computing, a brain-inspired method for dynamical systems modeling. We show that multi-year ENSO forecasts can be improved up to two years in advance. An important strength of our method is that it relies solely on past data, enabling realtime predictions without the need for future observations. In fine-tuning the parameters of our realtime filter, the Bayesian optimization is used for effectively capturing frequently occurring temporal patterns.
%We also use it to determine hyperparameters of our reservoir computing model. {\color{blue} (These two sentences sound as if we perform Bayesian optimizations twice for filter parameters and for reservoir parameters, respectively. But we optimize the patameters all together?)} 
%{\color{brown} The Bayesian optimizations for filter and RC model are performed separately in this work.} Our approach can also be applied to other complex, high-dimensional phenomena that demand robust and efficient long-term forecasting.\\
In this paper, we introduce a new type of filter that is applicable to realtime operational forecast and combine it with a machine-learning framework known as reservoir computing, a brain-inspired method for dynamical systems modeling. An important strength of our method is that it relies solely on past data, enabling realtime predictions without the need for future observations. The parameters of the realtime filter as well as the hyperparameters of the reservoir computing model are systematically calibrated with the Bayesian optimizations. Using this method, we show that the prediction horizon of ENSO can be extended up to two years. %In fine-tuning the parameters of our realtime filter, the Bayesian optimization is used for effectively capturing frequently occurring temporal patterns. We also use it to determine hyperparameters of our reservoir computing model. 
While we focus on ENSO forecasts in this article, our method can also be applied to a wide range of complex, high-dimensional phenomena that demand robust and efficient long-term forecasting.

\end{quotation}

\section{Introduction}\label{sec:introduction}
The El Ni\~{n}o-Southern Oscillation (ENSO) is the primary interannual fluctuation in the Earth's climate. It is characterized by the temperature anomaly in the central to eastern tropical Pacific Ocean. ENSO shows an oscillation among warm (El Ni\~{n}o), cold (La Ni\~{n}a), and neutral phases with a typical timescale of 3 to 8 years \cite{bjerknes_1969}. With the nature of teleconnection, climate variability has a strong effect on global atmospheric circulation, ecological systems, public health, and economics \cite{mcphaden2006}.

The prediction of the time evolution of ENSO has been continuously addressed throughout the past decades. The approaches include conceptual models \cite{schopf_1990, battisti_1989, wyrtki_1986, cane_1990}, statistical models \cite{trenberth_1996, grieger_1994}, and global climate models (GCMs)\cite{luo_2008,guilyardi2009}. The prediction using fully-coupled ocean-atmosphere models has demonstrated the high potential for ENSO prediction accuracy, yet challenges remain for long-term forecasts beyond a one-year lead time \cite{tang_2018}.

In recent years, as with the prediction of tropical cyclones \cite{loi_2024}, there has been a notable improvement in the prediction skills of ENSO using machine learning as well as using methods of complexity science \cite{ludescher2013improved,meng2020complexity}. There are historical examples of using neural networks to model nonlinear dynamics of ENSO \cite{grieger_1994, timmermann_2001, guardamagna2024detection}. In this century, it has become possible to achieve forecasts exceeding one year, with performance comparable to that of GCMs \cite{ham_2019,chen21, chen_2023, wang2024role}.

Reservoir computing is a brain-inspired machine-learning technique that employs a data-driven dynamical system \cite{Jaeger_2001,Jaeger_2004,Zhixin_2017,Lu_2018}. 
It is a method effective in predicting time series and frequency spectra in chaotic behaviors and is capable of learning time series of high-dimensional dynamical systems such as fluid flow ~\cite{nakai_2018,nakai_2020,kobayashi21,nakai_2024}. 
In particular, Nakai and Saiki~\cite{nakai_2020} clarified that delay-coordinates with appropriate delay time and dimension of the delay-coordinates are efficient when the number of observable variables is smaller than the effective dimension of the attractor. The reservoir computing can predict the statistical quantity %laminar lasting time distribution 
of a particular macroscopic variable of chaotic fluid flow, which
cannot be calculated from a direct numerical simulation of the Navier--Stokes equation because of its high computational cost~\cite{kobayashi21}. The reservoir computing approach has been found to be effective in prediction of complex atmospheric systems such as Madden-Julian Oscillation \cite{suematsu_2022}, the Asian summer monsoon \cite{mitsui_2021}, North Atlantic Oscillation \cite{Huang_2023} as well as ENSO \cite{hassanibesheli_2022,guardamagna2024detection}.

The aim of this study is to establish a general methodology for performing long-term prediction of various time series in operational realtime formulation, which excludes information from the future. 
In this study, applying the band-pass filter to the time series data is demonstrated to be effective for extending the prediction horizon of the ENSO time series. For a robust long-term prediction, we introduce a new data-driven filter.
It is known that high-frequency disturbances degrade the performance of machine learning prediction, while low-frequency behavior, such as trends and multidecadal variability, makes the construction of machine-learning models difficult \cite{hassanibesheli_2022}. 
In past studies, the formulation of predicting ENSO by artificial neural networks requires the application of some conventional bandpass filter techniques such as moving average\cite{guardamagna2024detection} or Butterworth filter \cite{hassanibesheli_2022,ham_2019}, which incorporate information from the future time steps into the filtered time series. Strictly speaking, even the conventional Ni\~{n}o indices include future information by the moving average. Therefore, those methodologies are not practical to be applied in a realtime operational forecast.
This study is the first successful attempt to develop a filtering method that extracts signals with cycles of 3 to 8 years using only past time series.
For the filtered Ni\~{n}o-3.4 time series data, we construct a data-driven model. 
In this study, we employ reservoir computing to construct a machine-learning model capable of predicting the multi-year dynamics of the ENSO for over two years.

%%%%%%%%%%%%%%%%%%%%%%%%%%%%%%%%%%%%%%%%%%%%%%%%%%
\section{Data and Methods}
\subsection{Data description}
We use monthly sea surface temperature (SST) for the 1870–2022 period, retrieved from the Hadley Centre Global Sea Ice and Sea Surface Temperature dataset \cite{hadisst}. The one-degree latitude-longitude grid data is averaged over the $5^\circ$N–$5^\circ$S, $170^\circ$W–$120^\circ$W regions in the tropical central and eastern Pacific, which are commonly used to define the Ni\~{n}o-3.4 index. To obtain the SST anomaly from the typical seasonal cycle, the 1971–2000 monthly climatological values are subtracted. Hereafter, we refer to this time series as the realtime SST anomaly. The conventional Ni\~{n}o-3.4 index uses 5-month running mean to define El Ni\~{n}o and La Ni\~{n}a events. However, in this study, we utilize monthly SST values without applying any moving average to exclude future information and enable realtime prediction. We then filter the unrequired frequency ranges by the realtime filter described in the next subsection.

\subsection{Realtime fitering method}
In this subsection, we present a new method to find a suitable filter for the corresponding time series data. Note that the corresponding Python codes of realtime filter and reservoir computing combined with Bayesian optimization are available in the repository \cite{Yoneda-code}. The pure reservoir computing code is also available.\cite{Tanaka-Nakane-Hirose}
To find the appropriate filter, we use Bayesian optimization 
and the following weight function of the moving average:
$$\Psi(t)=
 \underbrace{\left(d_1\cos\left(\frac{t}{\pi r_1}\right)+d_2\cos\left(\frac{t}{\pi r_2}\right)\right)}_{passing\ freq}
 \underbrace{\frac{(w-t)^c}{w^c}}_{band\ width}
 \quad\text{for}\quad t\in[0,w],$$
where 
$d_1,d_2,r_1,r_2,c,w>0$  are the parameters of the corresponding objective function \eqref{objective function}.
In what follows, we explain how to obtain this objective function.
Let $\Omega\subset\mathbb{Z}$ be a prescribed finite sequence.
Having obtained $\Psi$ by the optimization, we then applied the weighted moving average to the original data {$y:\Omega\to\mathbb{R}$} as follows:
\begin{equation*}
    \text{filtered data:}\quad y^*(t):=
    \sum_{t'=0}^{w}y(t-t')\Psi(t')
    \quad\text{for}\quad t\ 
    %\in\Omega
    %\quad\text{and}\quad
    \in\Omega.
    %\quad (t'=0,1,2,\cdots,w).
\end{equation*}
We emphasize that since the weight function $\Psi$ is localized in the timeline, in particular,
it does not contain any future information,
this bandwidth cannot have a unique threshold due to the Fourier uncertainty principle.
Here is a rough explanation of the bandwidth.
If $c$ is close to zero, then the Fourier transform of the corresponding weight function is close to the sinc function, which represents a narrow bandwidth (compared to the case when $c$ is large).
To the contrary, if $c$ tends to infinity, then 
the Fourier transform of the corresponding weight function tends to identically one (since $\Psi$ tends to the Dirac-delta function), which represents broad bandwidth (compared to the case when $c$ is small).
The main philosophy of this filter method is 
to effectively reduce the ``indefinite factor'' (such as white noise) without using the conventional Fourier transform, since the time domain of the training data $\Omega$ 
%{\color{blue}(or training data $\{y(t)|t\in \Omega\}$)}
is, of course, always finite, and, applying the Fourier transform to such finite elements is no longer mathematically rigorous.
To effectively reduce the indefinite factor, we use the idea of  
``dictionary'', in other word, ``key-value pairs''.
First, we discretize the range of the filtered data $y^*$ as follows:
For an integer $K$ which is greater than $1$,  we choose $\{a_k\}_{k=1}^{K}\subset\mathbb{R}$ such that 
%$a_k\not=a_{k'}$ ($k\not=k'$). 
$a_{1}<a_{2}<\cdots<a_K$. 
%$\sup\left\{\frac{a_{k+1}^K-a_{k-1}^K}{2}\ (1<k<K), \frac{a_{1}^K+a_{2}^K}{2}+1, 
%1-\frac{a_{K-1}^K+a_{K}^K}{2}\right\}\leq C/K$

\begin{figure}[H]
    \centering
    \includegraphics[width=0.8\columnwidth]{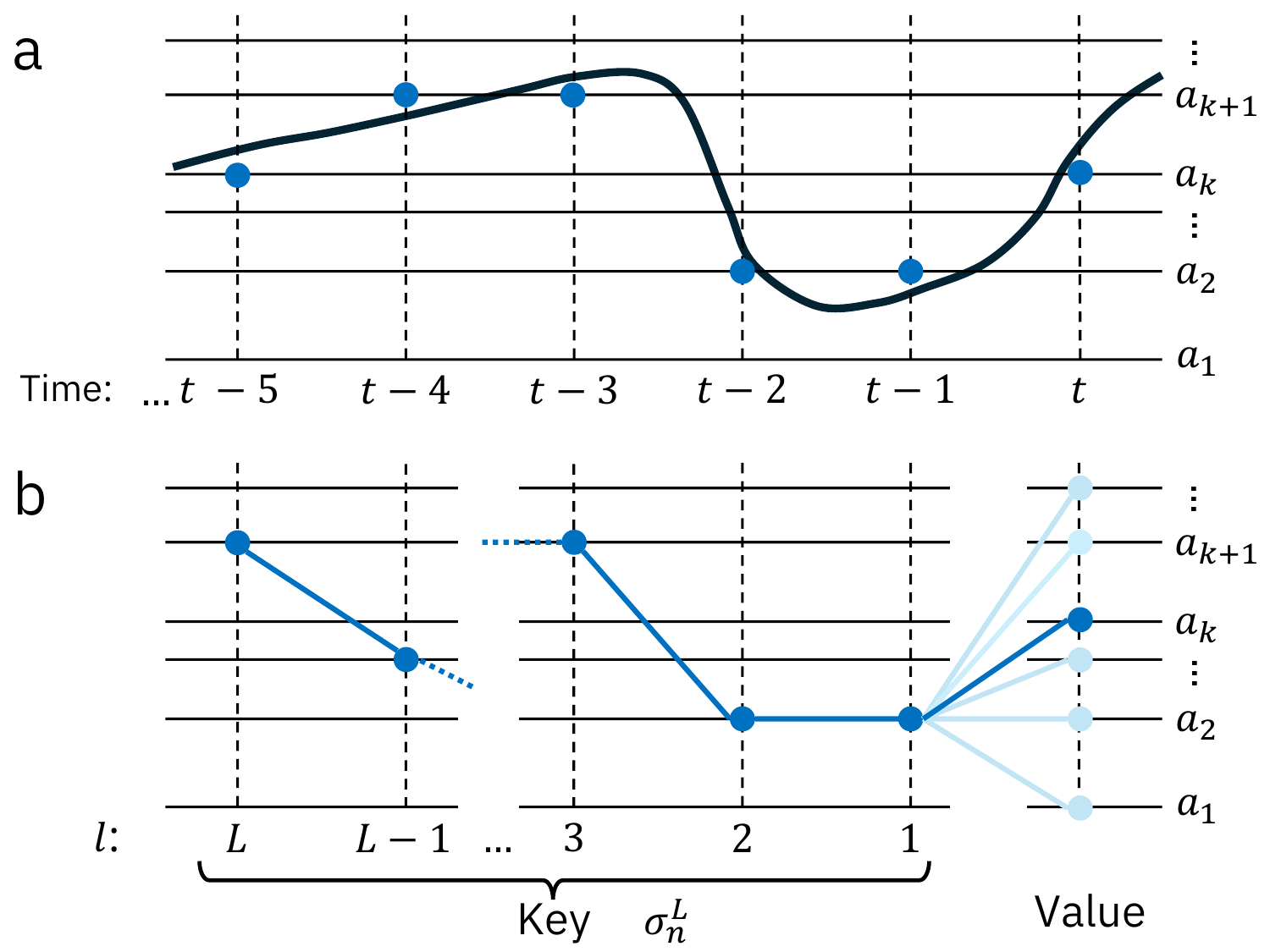} 
    \caption{{\bf Optimization of filter performances.} a) A filtered time series data $y^*$ (black thick line) is classified into discrete values $\tilde{y}$ (blue dots). b) Concept of the key and value: 
    %The indefinite factor of the filter is calculated statistically by the maximum rate at which a pattern $\sigma^L_n$ is followed by values $\{a_k\}^K_{k=1}$. 
    The indefinite factor of the filter is reduced statistically by maximizing \eqref{objective function}, which represents the highest rate at which a key pattern $\sigma^L_n$ is followed by one of the values $\{a_k\}^K_{k=1}$ at the subsequent time step.}
    \label{fig:dict}
\end{figure} 

\begin{remark}
   To determine the positions of $\{a_k\}_{k=1}^K$, we have applied the cumulative distribution function to $y^*$, more precisely,
   $\#\{t\in\Omega:a_k\leq y^*(t)<a_{k+1}\}$ ($k=1,2,\cdots,K-1$),  $\#\{t\in\Omega: y^*(t)<a_{1}\}$ and  $\#\{t\in\Omega:a_K\leq y^*(t)\}$ 
   to be almost the same integer.
\end{remark}
Using this $\{a_k\}_{k=1}^K$,  we classify the filtered data $y^*$ of each function value as follows:
\begin{equation}\label{discrete}
\tilde y(t):=\text{argmin}_{a\in\{a_k\}_{k=1}^{K}}|y^*(t)-a|\quad\text{for}\quad t\in\Omega.
\end{equation}
%where $a-0:=a-\varepsilon$ for any sufficiently small $\varepsilon>0$
See Fig.\ref{fig:dict} a.
To perform the pattern classification, we first choose the maximum and minimum length of the pattern a-priori:
$L_{max}$ and $L_{min}$ ($L_{max}>L_{min}\geq 1$), and let
%and then 
%we take $L=L_{max},L_{max}-1,\cdots, L_{min}+1,L_{min}$.
 $\sigma_n^L$ ($n=1,2,\cdots,N_L$) be a permutation operator (we say ``key'') such that
%, namely, a map
\begin{equation*}
\sigma_n^L:\{1,2,\cdots, L\}\to\{a_{1},a_{2},\cdots,a_{K-1},a_{K}\}
\qquad(\ell\mapsto \sigma_n^L(\ell))
\end{equation*}
for $L\in[L_{min},L_{max}]\cap\mathbb{Z}$ with $\sigma_n^L\not=\sigma_{n'}^L$ ($n\not=n'$).
Then there exists a suitable $N_L$ such that the permutation operators $\{\sigma^L_n\}_{n=1}^{N_L}$
%and the number $N_L$
%To find the number $N_L$, we just check
satisfy the following two properties:
\begin{equation}\label{all patterns}
\begin{cases}
\text{For any $t\in\Omega$, there is $n\in\{1,\cdots,N_L\}$ such that}\\
\qquad
\text{$\sigma_n^L(\ell)=\tilde y(t-\ell)$ for $\ell=1,2,\cdots, L$},\\
\text{For any  $n\in\{1,\cdots,N_L\}$ there is $t\in\Omega$ such that}\\
\qquad\text{$\sigma_n^L(\ell)=\tilde y(t-\ell)$ for $\ell=1,2,\cdots, L$.} 
\end{cases}
\end{equation}
Note that  $N_L\leq K^L$ due to the sequence with repetition.
%We now define the dictionary $\{(\sigma_n,a_{k(n)})\}_{n=1}^N$, in other word, key-value pairs (i.e. $\sigma_n$ is the ``key" and $a_{k(n)}$ is the corresponding ``value").
%We define the key value $a_{k(n)}^L$ and the corresponding variation as follows:
We now define sets of 
discrete-time $\Sigma_n^L\subset\Omega$ such that
\begin{equation*}
\Sigma_n^L:=\{t\in\Omega: \sigma_n^L(\ell)=\tilde y(t-\ell)\quad\text{for}\quad \ell=1,2,\cdots, L\}.
\end{equation*}
Then we can reduce the indefinite factor 
by using $\{\tilde y(t)\}_{t\in\Sigma_n^L}$ which is called ``value'', as displayed schematically in Fig.\ref{fig:dict} b).
%To be more precise, 
%let us define the mean $E_n^L$ and the corresponding variance $V_n^L$ as follows: 
%\begin{equation*}
%    E_n^L:=\frac{1}{|\Sigma_n^L|}\sum_{t\in\Sigma_n^L}\tilde y(t)
%    \quad\text{and}\quad
%    V_n^L:=\frac{1}{|\Sigma_n^L|}\sum_{t\in\Sigma_n^L}|\tilde y(t)-E_n|^2.
%\end{equation*}
%Now we use the Bayesian optimization to determine the appropriate parameters:\\ $r_1,r_2,d_1,d_2,c,w$.
%Under the condition that the autocorrelation between the original and the filterd data is not small,
% we require the following condition:
%\begin{center}
%the individual variances $V^L_n$ ($L\in [L_{min},L_{max}]\cap\mathbb{Z}, n\in [1,N_L]\cap\mathbb{Z}$) are small.
%\end{center}
%\begin{remark}
%However the above explanation (using variances) is only for simplicity.
%In the numerical computation, 
%we use the following nonparametric approach:
Let  
\begin{equation*}
    \Pi_n^L(k):=\{t\in\Omega: \sigma_n^L(\ell)=\tilde y(t-\ell)\quad\text{for}\quad \ell=1,2,\cdots,L \quad\text{and}\quad \tilde y(t)=a_k\},
\end{equation*}
and we maximize the following objective function %\eqref{objective function}  
under the condition that the autocorrelation between the original and the filtered data is not small:
\begin{equation}\label{objective function}
\displaystyle\frac{\displaystyle\sum_{L=L_{min}}^{L_{max}}\#\left\{n:\text{there is a }k\in\{1,2,\cdots,K\}
    \text{ such that } \frac{\#\Pi^L_n(k)}{\#\Sigma_n^L}\geq \gamma\right\}}{\displaystyle\sum_{L=L_{min}}^{L_{max}}N_L},
\end{equation}
where $\gamma\in(0,1]$ is the match rate of $a_k$ for each $\sigma_n^L$.
%\end{remark}
Here, $\gamma$, as well as $L_{max}$ and $L_{min}$ are prescribed parameters in the numerical implementation. $\gamma$ is close to 1. $L_{max}$ and $L_{min}$ are determined so that the length of frequent patterns is sufficiently covered. 
%{\color{blue}How $L_{max}$ and ${L_{min}}$ are set in the actual computation? What are the tuning variables of the obsective function(s)? Only $d_1,d_2,r_1,r_2,c,w>0$?}

\subsection{Reservoir computing}
The echo-state network (ESN) is a widely used reservoir computing architecture based on a recurrent neural network, consisting of an input layer, a reservoir state vector, and an output layer\cite{Jaeger_2001,Jaeger_2004,Zhixin_2017,Lu_2018}.
A vector of an observed variable as input and a reservoir state vector at time step $t$ is updated as 
\begin{equation}
\begin{cases}
  \mb{r}(t+\Delta t) = (1-\alpha)\mb{r}(t)+\alpha \tanh(\mb{A}\mb{r}(t) + \sigma_{\text{in}}\mb{W}_{\text{in}}\mb{u}(t)) \\
  \mb{u}(t+\Delta t) = \mb{W}_{\text{out}}\mb{r}(t+\Delta t),
  \label{eq:reservoir}
\end{cases}
\end{equation}
where 
$\mb{u}(t) \in \mathbb{R}^M$ is the input vector of the time series data, 
$\mb{r}(t) \in \mathbb{R}^N~(N \gg M)$ is the reservoir state vector, 
$\Delta t$ represents the time interval at which the time series is updated (1 month in this study), and 
$\sigma_\text{in}$ is the scaling parameter of the input matrix %$W_{\text{in}}$
$\mb{W_{\text{in}}}$.
$\mb{W}_{\text{in}} \in \mathbb{R}^{N\times M}$ 
and $\mb{W}_{\text{out}} \in \mathbb{R}^{M\times N}$
are input and output matrices;
$\mb{A} \in \mathbb{R}^{N\times N}$ is a matrix that depicts the recurrent connectivity within the reservoir. The density of the non-zero elements of $\mb{A}$ is determined by the parameter $p$.
The maximum absolute value of the eigenvalue of $\mb{A}$ is called the spectral radius and denoted by $\rho$.
The spectral radius is related to the echo state property (ESP), which means that the effect of initial conditions vanishes after a certain amount of time. Setting the spectral radius to a value less than unity has been used as a rule of thumb to ensure that the model possesses the ESP. However, ESP often holds for $\rho > 1$ \cite{Luk2012}. Yildiz et al. (2012) demonstrated that no universally applicable rule exists between the ESP and the optimal spectral radius\cite{yildiz2012}. It should be determined through task-specific experimentation, taking into account the nature of the driving input. In the present study, we impose the constraint $\rho < 1$ with the intention of limiting the parameter search space in Bayesian optimization, thereby facilitating the identification of the optimal value.
$\alpha$ ($0<\alpha\le 1$) is the coefficient called leaking rate that adjusts the smoothness of the time evolution of $\mb{r}(t)$. 
Hyperbolic tangent function for a vector $\mb{q} = (q_1,q_2,\ldots,q_N)^{\text{T}}$ is defined as $\tanh(\mb{q})=(\tanh(q_1), \tanh(q_2),\ldots,\tanh(q_N))^{\text{T}}$ 
where superscript $\text{T}$ represents the transpose.

\begin{figure}[H]
    \centering
    \includegraphics[width=0.9\columnwidth]{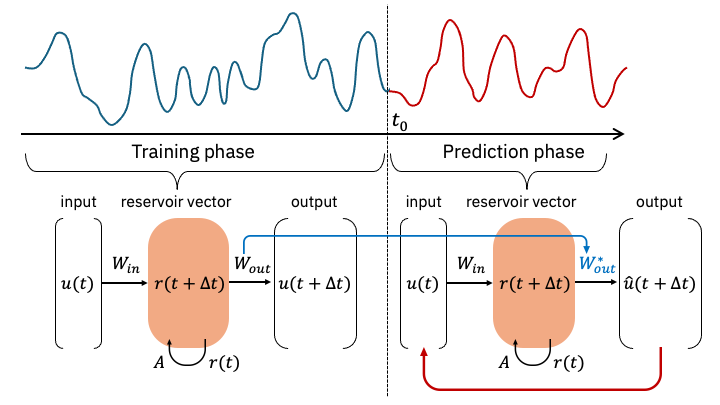} 
    \caption{{\bf Schematic diagram of reservoir computing.} 
    In training phase ($t<t_0$), optimal value of $\mb{W}_{\text{out}}$ is determined using the input time series with prescribed random matrix $\mb{A}$ and $\mb{W}_{\text{in}}$. In the prediction phase ($t\ge t_0$), the dynamics of the filtered time series are inferred recursively by the optimized model of the output matrix, which is denoted as $\mb{W}^*_{\text{out}}$.}

    \label{fig:reservoir}
\end{figure} 

The choice of input and output variables of the reservoir computing and their dimensions is essential for constructing the modeling. 
It is known that a chaotic attractor can be reconstructed by some observable variable and its delay-coordinates~\cite{takens_1981, sauer_1991non}.
For the reservoir computing models, it has been reported that a delay-coordinate variable with appropriate delay time is efficient when the number of observable variables is smaller than the effective dimension of the  attractor~\cite{nakai_2020,huang_2020,Huang_2023}. Thus, we employ the choice of delay coordinates of an observable variable as a hyperparameter of the model.

Let $y^*$ be the filtered realtime SST anomaly.
For these reasons above, we introduce an $M$-dimensional delay-coordinate vector of filtered realtime SST anomaly with a delay-time $\Delta \tau$ as input and output data $\mathbf{u}$ in Eq.~\eqref{eq:reservoir}:
\begin{align*}
    \mathbf{u}(t)=(y^*(t),
                    y^*(t-\Delta \tau),
                    \ldots,
                    y^*(t-(M-1)\Delta \tau))^T,
\end{align*}
where the optimal values of $(\Delta \tau, M)$ are determined using the Bayesian optimization framework Optuna \cite{optuna_2019}, along with other hyperparameters as described in the next section. For each set of training and prediction, let $t_0$ be the initial time step of prediction. The preceding 100-year period of the data $\mb{u}(t)\ (t_0 - 1200 \leq t \leq t_0-1)$ is used as training data.

In the training phase, with a fixed pair of matrices $\mb{A}$ and $\mb{W}_{\text{in}}$, 
$\mb{W}^*_\text{out}$ is optimized using training data so that  
$\mb{W}^*_\text{out}\mb{r}(t)\approx\mb{u}(t)$ is satisfied (Fig.~\ref{fig:reservoir} left). Here the cost function 
\begin{equation}
   \Lambda = \sum^{T_\text{train}}_{t=1}||\mb{W}_{\text{out}}\mb{r}(t)-\mb{u}(t)|| + \beta||\mb{W}_{\text{out}}||^2
\end{equation}
is minimized to obtain the best inference of the model, where $T_\text{train}$ is the length of training data and $||..||$ is $L_2$-norm of a vector. The optimal output matrix is 
\begin{equation}
  \mb{W}^*_{\text{out}} = \delta\mb{U}\delta\mb{R}^T(\delta\mb{R}\delta\mb{R}^T + \beta \mb{I})^{-1}
\end{equation}
where the columns of the matrices $\delta\mb{R}$ and $\delta\mb{U}$ corresponds to the vectors $\mb{r}(t)$ and $\mb{u}(t)$, respectively, in chronological order. $\mb{I}$ denotes identity matrix. %{\color{blue} How is the constant bias implemented? Something like $\mb{u}(t)=\mb{W}_\text{out}r(t)+\mb{c}$.} 

In the prediction phase, the output matrix is fixed to the optimal value $\mb{W}^*_\text{out}$ obtained (Fig.~\ref{fig:reservoir} right). We then get the predicted value $\hat{\mb{u}}(t+\Delta t)$ from $\mb{u}(t)$ and $\mb{r}(t)$, using Eq.~\eqref{eq:reservoir} with the same $\mb{A}$ and $\mb{W}_{\text{in}}$. The output vector is fed back as input to predict the value at the next time step $t+2\Delta t$. In this manner, the time integration is recursively advanced to obtain the time evolution up to 36 months ahead.

\subsection{Bayesian optimization of hyperparameters}
For both the realtime filter and the reservoir model, the combinations of hyperparameters are selected using Bayesian optimization with Optuna \cite{optuna_2019}.
The hyperparameters of the realtime filter are chosen to maximize the product of the objective function in Eq.~(\ref{objective function}) when applied to the realtime SST anomaly, and the maximum lag-correlation between the original and filtered monthly SST anomaly.
The hyperparameters of the reservoir computing model, namely, $M$, $\Delta \tau$, $N$, $\beta$, $p$, $\rho$, $\alpha$, and $\sigma_\text{in}$, as well as the random seed to generate the matrices $\mb{A}$ and $\mb{W_{\text{in}}}$, are chosen based on the prediction performance of 120 sequences of training and predictions whose $t_0$ ranges from January 1986 to December 1995.
%{\color{blue}What is the performance here? Are the filter performance (the factor in Remark II.2) and the prediction performance of resevoir maximized together? If so, are there two objective functions to be minimized?} 
 As a measure of the model performance, we use the all-season correlation skill $C(\mu)$ 
\begin{equation}
    C(\mu) = \frac{1}{12}\sum_{m=1}^{12} \frac{\sum_{\eta=\eta_s}^{\eta_e}(Y_{\eta,m}-\overline{Y_m})(P_{\eta,m,\mu}-\overline{P_{m,\mu}})}{\sqrt{\sum_{\eta=\eta_s}^{\eta_e}(Y_{\eta,m}-\overline{Y_m})^2\sum_{\eta=\eta_s}^{\eta_e}(P_{\eta,m,\mu}-\overline{P_{m,\mu}})^2}}
    \label{eq_corr}
\end{equation}
as a function of the forecast lead months $\mu$. Here, $Y$ and $P$ denote the observed and the predicted values, respectively. $\overline{Y_m}$ and $\overline{P_{m,l}}$ denote the climatologies with respect to the calendar month $m$ (from 1 to 12) and the forecast lead months $\mu$. The label $\eta$ denotes the forecast target year. $\eta_s$ and $\eta_e$ denote the earliest (that is, 1986) and the latest year (that is, 1995) of the period, respectively. $C(\mu=24)$ is maximized here.
The determination of the random matrix $\mb{A}$ is as follows. First, using two consecutive integers $i$ and $i+1$ as seed values, two $N\times N$ matrices $\mb{A}_i$ and $\mb{A}_{i+1}$ are generated with a density of non-zero elements equal to $p$. Next, by linearly interpolating between the two matrices while maintaining the sparsity $p$, a matrix $\mb{A}_{i+x}$ characterized by a real seed value $x$ $(0<x<1)$ is created. Finally, $\mb{A}_{i+x}$ is rescaled to match the condition of spectral radius $\rho$.
For the determination of the random matrix $\mb{W}_{\text{in}}$, it is similarly defined as a matrix obtained by linearly interpolating between two random matrices $\mb{W}_{i}$ and $\mb{W}_{i+1}$.

%%%%%%%%%%%%%%%%%%%%%%%%%%%%%%%%%%%%%%%%%%%%%%%%%%
\section{Result}
\subsection{Performance of the new filter}
\begin{figure}[H]
\centering	
            \subfigure[]{%
		\includegraphics[width=0.40\columnwidth]{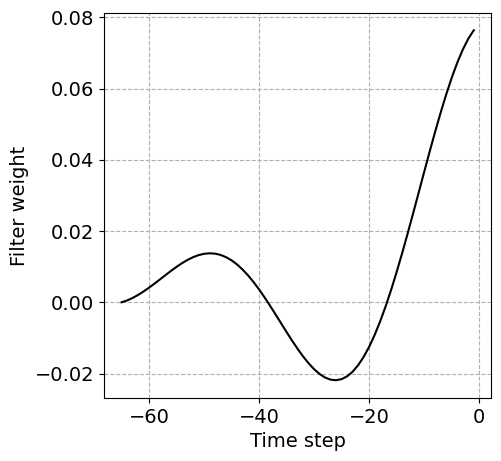} }
            \subfigure[]{%
		\includegraphics[width=0.38\columnwidth]{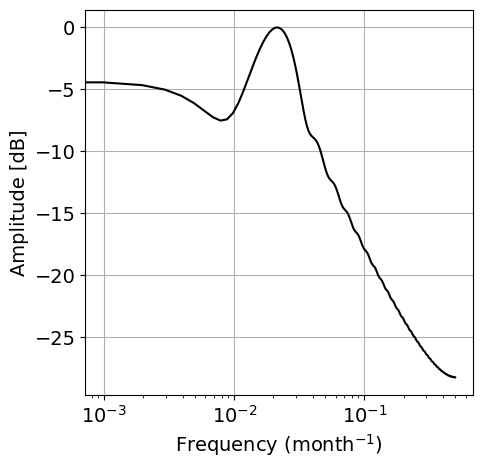} }
		\caption{{\bf Characteristics of the new realtime filter.} (a) The weight function in terms of time step and (b) frequency response function of the weight function. The weight function for $t \ge 0$ is zero so that the filtered time series does not include information from the future. The left endpoint of the weight function is also determined by Optuna in the parameter $w$. %used in the convolution is determined by identifying the time at which its amplitude becomes sufficiently small. 
        %In this case, we use the interval $-65\le t < 0$.
        }
		\label{fig:char_filter}
	\end{figure}

As a result of the Bayesian optimization, the best set of filter parameters is obtained as shown in Table \ref{param_filter}. The shape of the filter function is displayed in Fig.\ref{fig:char_filter}a. The weight equals zero everywhere $t\geq 0$. The frequency response of the filter. The filter is applied by performing a convolution integral of the weighting function shown in Fig.\ref{fig:char_filter}a with the original time series. Fig.\ref{fig:char_filter}b shows that the passband of the filter lies in about four to eight-year periods, which is roughly equal to the typical timescale of the perturbation of ENSO.

\renewcommand{\arraystretch}{0.7}
\begin{table}[H]
 \begin{center}
  \begin{tabular}{cl} \hline
    Parameter & Value \\ \hline
    $r_1$ & 39.333\\
    $r_2$ & 2.789 \\
    $d_1$ & 0.152 \\
    $d_2$ & 0.448 \\
    $c$   & 1.086 \\
    $w$   & 65 \\
  \end{tabular}
  \end{center}
  \caption{The list of parameters and their values for the realtime filter obtained from the Bayesian optimization.}
  \label{param_filter}
\end{table}

\begin{figure}[H]
\centering	
            \subfigure[]{%
		\includegraphics[width=0.92\columnwidth]{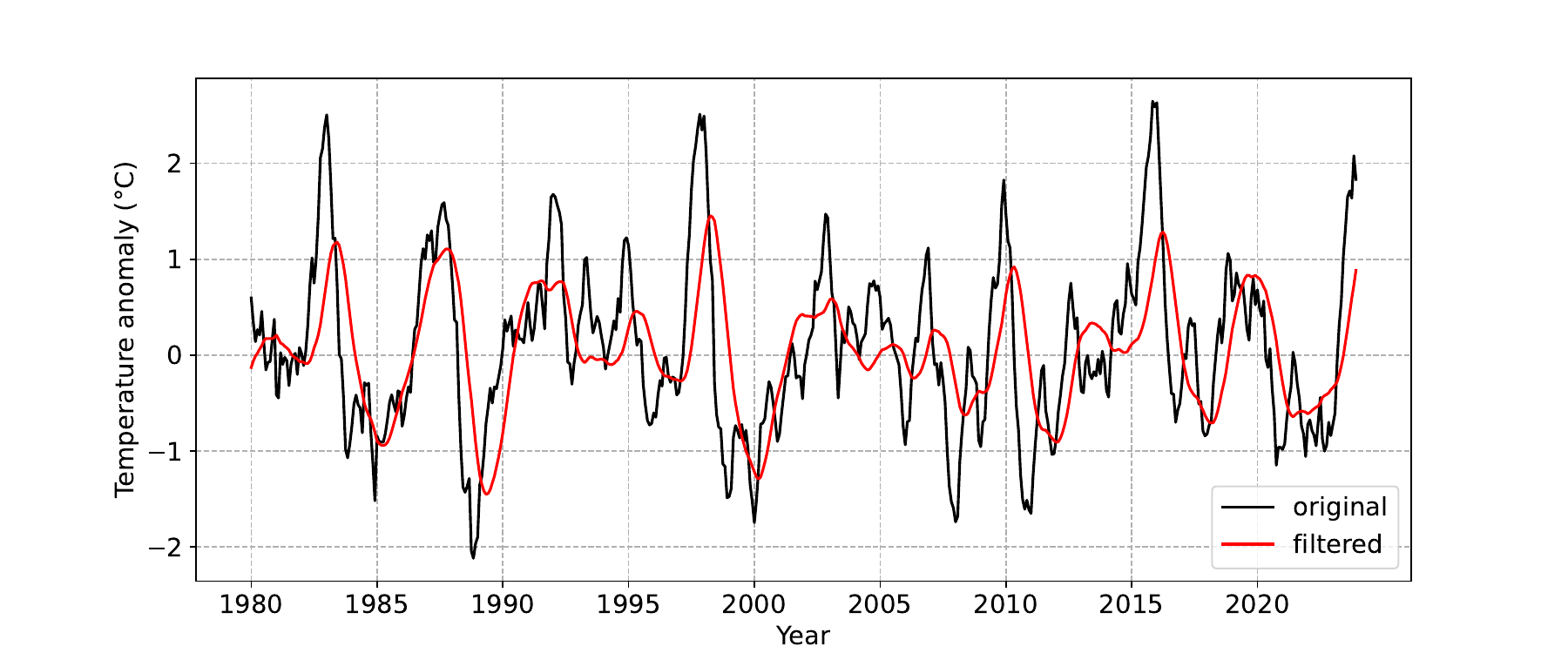} }
            \subfigure[]{%
		\includegraphics[width=0.75\columnwidth]{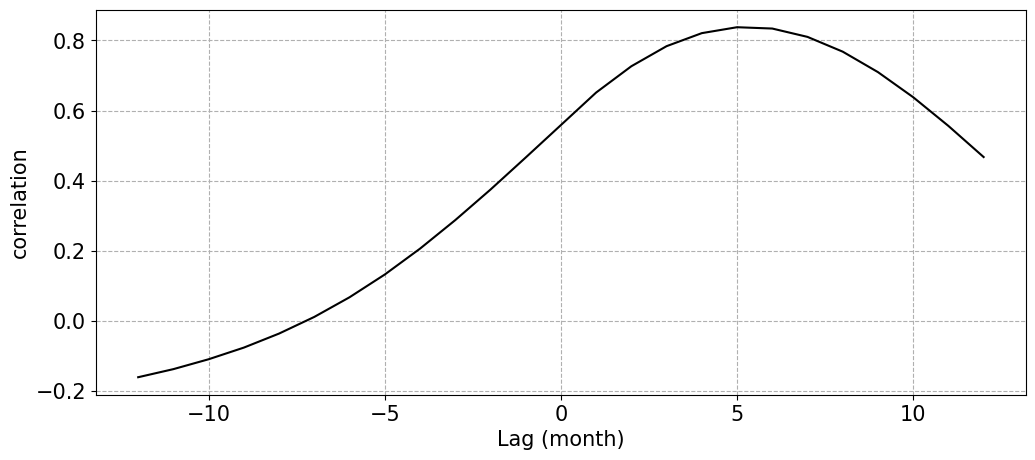} }
		\caption{{\bf Comparison between the original and filtered realtime SST anomaly.} (a) The time series of original (black) and filtered (red) indices from 1980 to 2022. (b) Lag correlation between the two time series over the entire data period. When the lag $\lambda$ is positive, it corresponds to delaying the original time series by $\lambda$ steps.}
		\label{fig:comparison_filter}
	\end{figure}

By applying the new realtime filter, we obtain a target time series that preserves information on long-term ENSO variations while enhancing the predictive performance of the reservoir computing model. The filtered realtime SST anomaly is shown 
in Fig.\ref{fig:comparison_filter}a along with the original time series. Due to the asymmetry of the filter function with respect to the positive and negative time ranges, the overall time series shifts backward by approximately five months. The shift is highlighted by the lag correlation between the original and the filtered time series (Fig.\ref{fig:comparison_filter}b). The maximum correlation is 0.837 with a lag of five months, and we assume that the filtered realtime SST anomaly can be used as an alternative index for the state of ENSO.

\subsection{Prediction skill of the model}
\label{sec:predictionskill}
As a result of Bayesian optimization, the optimal set of hyperparameters for the reservoir computing model is obtained as listed in Table \ref{param_model}. 
To evaluate the prediction skill of the current method, the sequence of training and prediction is conducted 180 times with different $t_0$ every month between January 2001 and December 2015. The prediction skill of the filtered realtime SST anomaly was evaluated using the all-season correlation skill in Eq.~(\ref{eq_corr}) with $\eta_s=2001$ and $\eta_e=2015$.

\renewcommand{\arraystretch}{0.7}
\begin{table}[H]
 \begin{center}
  \begin{tabular}{cl} \hline
    Parameter & Value \\ \hline
    $\Delta \tau$ & 4 \\
    $M$ & 9 \\
    $N$ & 244 \\
    $\beta$ & 0.759 \\
    $p$   & 0.290 \\
    $\sigma_\text{in}$   & 0.477 \\
    $\rho$ & 0.712 \\ 
    $\alpha$ & 0.975 \\
  \end{tabular}
  \end{center}
  \caption{The list of parameters and their values for the reservoir computing model obtained from the Bayesian optimization.}
  \label{param_model}
\end{table}

\begin{figure}[H]
\centering	
		\includegraphics[width=0.7\columnwidth]{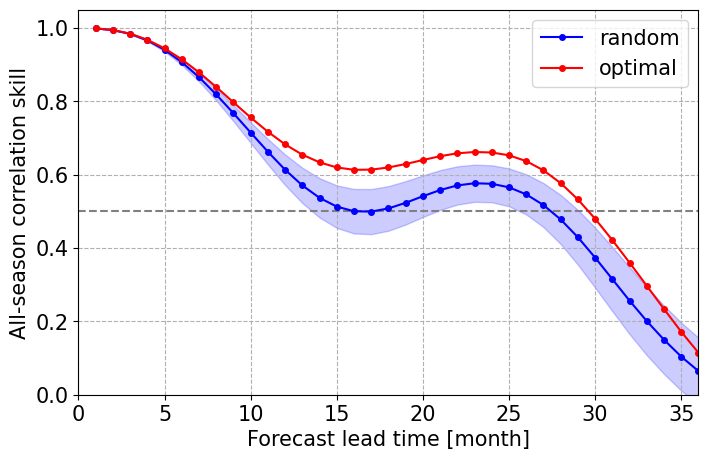} 
		\caption{{\bf Prediction skill of the data-driven model. } All-season correlation skill for ecch forecast lead time is obtained from 180 predictions between January 2001 and December 2015 based on Eq. (\ref{eq_corr}). The red line shows the prediction obtained using the optimal matrices $\mb{W}_{\text{in}}$ and $\mb{A}$, which were determined simultaneously with the hyperparameters during the model selection process. The blue solid line represents the average of the correlation skill when randomly generated 100 combination of $\mb{W}_{\text{in}}$ and $\mb{A}$ are used with the same hyperparameters as in Table \ref{param_model}, while the blue shading indicates the range of ±1 standard deviation. The horizontal gray dashed line denotes a correlation of 0.5, displayed as a benchmark for the model’s prediction capability.}
		\label{fig:corr_skill}
	\end{figure}

We focus on enhancing our data-driven model’s ability to predict the long-term behavior of ENSO over timescales beyond one year. 
The definition of successful prediction is based on maintaining a correlation skill above 0.5 for a given period, which is a criterion commonly used in previous studies\cite{hassanibesheli_2022, ham_2019, wang2024role}.
With the optimal hyper-parameter settings of the realtime filter and the reservoir computing model, the average all-season correlation skill remains above 0.5 for 29 months (red line in Fig. \ref{fig:corr_skill}). 
Even discounting the fact that the filtered time series to be modeled is shifted about 5 months into the future due to the asymmetry of the filter function, the performance of the current model succeeds in predicting the dynamics of ENSO for 24 months (i.e., two years). 
Note that in this case, the predictions are made using a fixed set of $\mb{A}$ and $\mb{W_{\text{in}}}$ that maximizes the objective function during the model selection process. In Fig. \ref{fig:corr_skill}, the correlation skill of the same evaluation period is also displayed for the cases where 100 sets of $\mb{A}$ and $\mb{W_{\text{in}}}$ are randomly selected (blue solid line for the mean and shading for the standard deviation). From this figure, it can be observed that the optimization in the model selection process allows for extracting optimal random matrices that result in a reservoir model with outstanding prediction skill in the model evaluation process over a different data period.

The main achievement of this study is the machine learning prediction of the long-term dynamics of ENSO, with a lead time up to two years, strictly avoiding the use of future information by the realtime filter. 
The two-year prediction horizon is achieved using the simpler architechture of reservoir computing, compared to the deep neural network model by Wang et al. (2024)\cite{wang2024role}, while it should be noted that the target of prediction in this study is the filtered time series.

\begin{figure}[H]
\centering	
		\includegraphics[width=0.85\columnwidth]{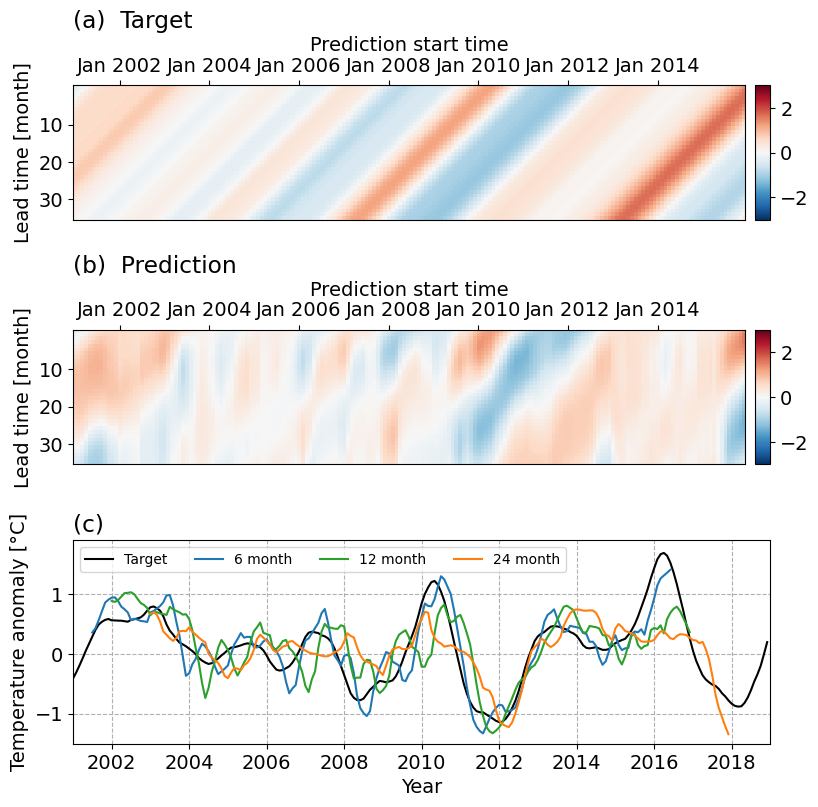} 
		\caption{{\bf Comparison between the predicted and the target time series of filtered realtime SST anomaly.} (a) The reference target and (b) the 180 runs of prediction. Each column in the two-dimensional plot corresponds to an individual run of 36-month prediction. The vertical axis indicates the lead time from the start of the prediction, while the horizontal axis shows the prediction start time step, with January 2001 set as zero. (c) Comparison of the prediction versus the target filtered realtime SST anomaly for different lead months: 6 months (blue line), 12 months (green line), and 24 months (orange line).}
		\label{fig:pred_ref}
	\end{figure} 

The predicted time series (Fig. \ref{fig:pred_ref}b) are displayed along with the corresponding target (Fig. \ref{fig:pred_ref}a). In some successful cases (e.g. those that start during 2010 and 2011), we have been able to capture the major positive and negative peaks of the filtered realtime SST anomaly up to about three years in advance. The predictions for different lead times (Fig. \ref{fig:pred_ref}c) show a high visual similarity to the target (filtered realtime SST anomaly). In particular, even the prediction with a lead time of 24 months faithfully captures the large-amplitude cycle of ENSO observed from 2010 to 2014. It should also be noted that the long-term predictions of the model in this study show relatively low accuracy in representing the signal of the record-breaking strong El Ni\~{n}o that occurred between 2015 and 2016.

%%%%%%%%%%%%%%%%%%%%%%%%%%%%%%%%%%%%%%%%%%%%%%%%%%
\section{Discussions and conclusions}\label{sec:conclusion}
\subsection{Discussions}
\begin{figure}[H]
\centering	
		\subfigure[]{
            \includegraphics[width=0.48\columnwidth]{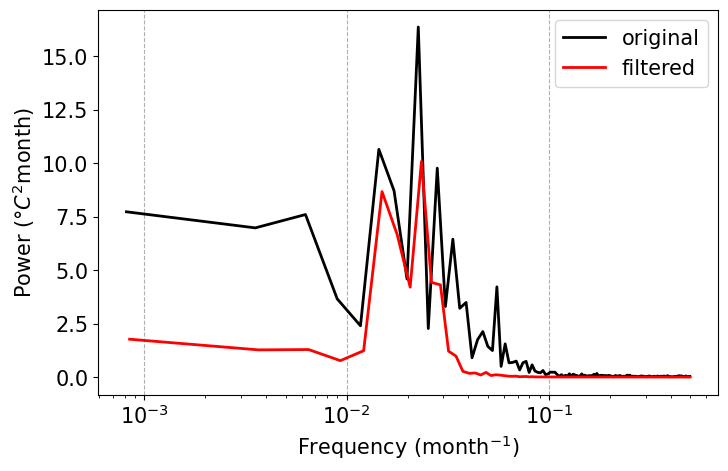} }
            \subfigure[]{
            \includegraphics[width=0.48\columnwidth]{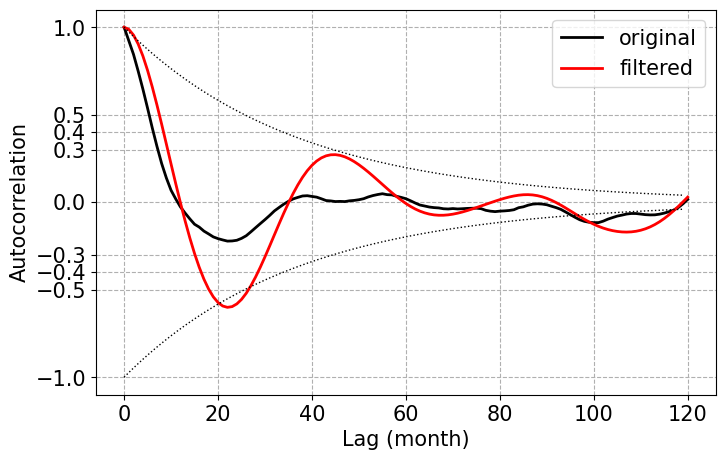} }
		\caption{{\bf Temporal characteristics of the original and filtered realtime SST anomaly.} (a) Power spectrum of the original (black line) and filtered (red line) time series. (b) Autocorrelation of the time series. Black dotted lines denote visually estimated envelope of filtered time series ($\pm\exp[-t/37]$).}
		\label{fig:power}
	\end{figure}

In this study, we used a delay-coordinate vector of filtered realtime SST anomaly as input and output of the ESN model. According to the model construction guidelines provided by Nakai and Saiki (2021)\cite{nakai_2020}, setting the value of the delay $\Delta\tau$ such that the consecutive two components of a delay-coordinate vector have a certain degree of correlation allows for capturing the fast dynamics. In addition, $\Delta\tau M$ should be chosen so that the envelope of the autocorrelation of the target time series is between 0.35 and 0.4 to capture the slow macroscopic dynamics. The envelope estimated from the filtered realtime SST anomaly used in this study (dotted line in Figure \ref{fig:power}b) passes through this range between lags of 35 to 39 months. Therefore, the hyperparameter combination obtained through the Bayesian optimization (Table \ref{param_model}, $\Delta\tau M=36$) meets this criterion. By incorporating past information from up to 36 months before the forecast start time, it is inferred that the long-term dynamical information of ENSO is effectively captured.

In our estimation, the dimension of the reservoir state vector is $N=244$, and for several other good combinations of hyperparameters obtained through Bayesian optimization, $N$ typically ranges from 100 to 300. One limitation of our current approach is that, considering the long-term dynamics of ENSO of interest, the available length of time series data based on historical observations as training data is limited. In the current work, setting a large number of nodes in the reservoir does not offer a significant advantage in reproducing long-term dynamics. If longer data are obtained by some data augmentation, increasing the number of nodes may allow for constructing models that can express more complex dynamics.

In complex phenomena, such as atmosphere-ocean coupled dynamics, multiple processes operating at various speeds are intertwined, forming cross-scale interactions. In such cases, it is known that the prediction accuracy of the reservoir computing model varies depending on the passband of the filter\cite{hassanibesheli_2022,huang_2020}. Figure \ref{fig:power} shows the power spectrum of the time series before and after applying the filter. It can be observed that the filter significantly reduces fluctuations with periods shorter than approximately three years and those longer than approximately eight years. For a practical methodology for prediction, it is necessary to identify filter parameter settings that maximize the model's predictive performance while minimizing the deviation of the filtered time series from its original.

In this paper, fluctuations with periods shorter than one to two years are removed. Here, the same metric as in previous research is used for prediction skill, but the target time series is the filtered realtime SST anomaly, which is not identical to the conventional Ni\~{n}o 3.4 index. Therefore, a direct comparison of the prediction horizon to those of previous studies is not possible. However, since the correlation between the original and filtered time series is high, it can be said that similar time series are being predicted. When trying to accurately predict the most unstable high-frequency dynamics, the accuracy of long-term predictions tends to decrease. Thus, predicting the original realtime SST anomaly is not the aim of this study. However, considering that the all-season correlation skill gradually decreases from 1 in Fig. \ref{fig:corr_skill} and the high correlation between the original and filtered time series,  it can be inferred that the reliability of short-term predictions for several months is not significantly compromised.

Furthermore, from the prediction skill, it is possible to gain some insights into the time scales of ENSO dynamics that each model is particularly proficient in forecasting. In the model of Hassanibesheli et al. (2022)\cite{hassanibesheli_2022}, the correlation remains at a high value of around 0.8 until 14 months of lead time. The advantage of combining low-frequency variation prediction using an ESN with high-frequency variation prediction using a statistical method called ``past noise forecasting'' is reflected in the prediction skill at a timescale of approximately one year. 
In this study, it is suggested that the reservoir computing model can more easily capture the primary timescale variations of ENSO through this filtering process. By applying the filter, the local minimum at about 22 months and the local maximum at about 44 months of the autocorrelation of the realtime SST anomaly are amplified (Figure \ref{fig:power}b). This implies that the periodic fluctuations of about 44 months are relatively enhanced, which in turn improves the model’s predictive performance on the lead time of half of the period, namely, approximately 22 months. 

One characteristic observed in the all-season correlation skill of the reservoir model’s predictions is that, it decreases from unity until reaching a local minimum at around 15 months, and then increases again up to a lead time of 23 months (Figure \ref{fig:corr_skill}). Similar temporary increase in correlation or decrease in relative error have been reported in previous studies\cite{hassanibesheli_2022,Huang_2023}. 
The non-monotonic behavior in prediction skill is presumably due to the calculation method of the metric. As shown in Eq. (\ref{eq_corr}), on calculating $C(\mu)$, only the value at the prediction lead time $\mu$ is used, rather than using the time series from lead time 1 to $\mu$. Therefore, it is possible for longer-term predictions to have higher scores than shorter-term predictions.
A more detailed investigation into the relationship between the prediction skills and the passband settings of the filter is required in our future study.

Determining the optimal combination of hyperparameters for machine learning using a simple grid search becomes increasingly difficult as the number of hyperparameters grows.
Considering the simplicity of the architecture and the large number of hyperparameters in reservoir computing, Bayesian optimization for tuning is an effective solution for improving its prediction performance\cite{joy_2022,optuna_2019}.
In this study, six parameters were tuned for the optimization of the filter, and nine parameters were tuned for the optimization of the reservoir model. In both cases, up to 1000 combination trials were conducted. The final performance did not change after about 100 trials for the filter, and after about 200 trials for the reservoir model. Therefore, we believe that the high dimensionality of the search space is not a crucial barrier for the specific task in this study. With the advancement of state-of-the-art optimization algorithms, it is expected that future research will be able to utilize more sophisticated reservoir models with an increased number of hyperparameters.

%\allblack
%{\color{orange} TM: The correlation skill score of Hassanibesheli et al. (2022) exhibits a similar local maximum at 12 months. This might also indicate annual nature of ENSO dynamics, though not bi-annual in their case. BTW the autocorrelation of Niño-3.4 index has a negative peak at around 24 months (again Hassanibesheli et al. 2022). Is this aspect related to our second peak at 23 months?}

\subsection{Conclusions}
In this study, using ENSO as an example, we proposed a machine learning methodology that enables long-term prediction of time series in complex phenomena characterized by high-dimensional dynamical systems. By applying a realtime filter that passes temporal scales characteristic of the target time series, and using delayed coordinates of the filtered time series as input variables to a reservoir computing model, we demonstrated improvements in performance of our prediction. A notable feature of this method is that it can be carried out from time series data generation to the pre-forecast process without incorporating any future information, thereby enabling operational realtime prediction. Since the workflow incorporates a Bayesian estimation algorithm to discover the optimal combination of hyperparameters for both the filter and the data-driven model, it suggests that this approach can be applied to dynamical phenomena other than ENSO as well.

\begin{acknowledgments}
Research of  TY  was partly supported by the JSPS Grants-in-Aid for Scientific
Research 24H00186.
TM was supported by the JST Moonshot R\&D (JPMJMS2389).
KN was supported by JSPS KAKENHI Grant No.22K17965. 
YS was supported by JSPS KAKENHI Grant No.19KK0067, 21K18584, 23H04465 and 24K00537.
\end{acknowledgments}

\section*{Author Declarations}
\subsection*{Conflict of Interest}
The authors have no conflicts to disclose.

\subsection*{Author Contributions}
\textbf{Takuya Jinno:} Conceptualization (lead); Data Curation (lead); Formal analysis (lead); Software (equal); Validation (equal); Visualization (lead); Writing - original draft (equal). \textbf{Takahito Mitsui:} Data Curation (supporting); Methodology (supporting); Writing - original draft (supporting). \textbf{Kengo Nakai:} Methodology (lead); Software (supporting); Writing - original draft (supporting). \textbf{Yoshitaka Saiki:} Funding (supporting); Methodology (supporting); Writing - original draft (supporting). \textbf{Tsuyoshi Yoneda:} Conceptualization (supporting); Formal analysis (supporting); Funding acquisition (lead); Software (equal); Validation (equal); Writing - original draft (equal).

\section*{Data Availability Statement}
The data that support the findings of
this study are available from the
corresponding author upon reasonable
request.

% If in two-column mode, this environment will change to single-column format so that long equations can be displayed. 
%\begin{widetext}
%$$\mbox{put long equation here}$$
%\end{widetext}

% Figures should be put into the text as floats. 
% Use the graphics or graphicx packages (distributed with LaTeX2e).
% Put the label that you will use with \ref{} command in the braces of the \label{} command.
%
% \begin{figure}
% \end{figure}

% \begin{table}
% \end{tabular}
% \end{table}

%\begin{acknowledgments}
% Put your acknowledgments here.
%\end{acknowledgments}

%\bibliography{reservoir_bibliography,qr_bibliography,heterochaos_bibliography.bib,mjo_reservoir.bib, enso_reservoir.bib,enso_reservoir_mitsui.bib}
\bibliography{enso_reservoir}
\end{document}